\documentstyle[spie]{article}  
\input{psfig}

\def\la{\lower 2pt \hbox{$\;\scriptscriptstyle \buildrel<\over\sim\;$}} 
\def\ga{\lower 2pt \hbox{$\;\scriptscriptstyle \buildrel>\over\sim\;$}} 
\def\lapp{\ifmmode\stackrel{<}{_{\sim}}\else$\stackrel{<}{_{\sim}}$\fi}
\def\gapp{\ifmmode\stackrel{>}{_{\sim}}\else$\stackrel{>}{_{\sim}}$\fi}

\def\sax{{\it BeppoSAX\ } }
\def\rxte{{\it RXTE\ }}

\def\etal{{\it et al.\ }}

\def\apj{{\em Ap. J.\ }}
\title{Use of Gas Electron Multiplier (GEM) Detectors for an Advanced X-ray Monitor}  
 
\author{Ronald A. Remillard\supit{a}, Alan M. Levine\supit{a}, Edward A. Boughan\supit{a}, Hale V. Bradt\supit{a}, \\ Edward H. Morgan\supit{a}, Ulrich J. Becker\supit{b}, Seppo Nenonen\supit{c}, and Osmi R. Vilhu\supit{d} 
\skiplinehalf  
\supit{a}Center for Space Research, Massachusetts Institute of Technology, \\
Cambridge, MA, USA \\
\supit{b}Laboratory for Nuclear Science, Massachusetts Institute of Technology, \\ Cambridge, MA, USA \\
\supit{c}Metorex International Inc., Helsinki, Finland \\
\supit{d}Observatory, University of Helsinki, Helsinki, Finland }
 
\authorinfo{Further author information: Send correspondence to R.A.R.; E-mail: rr@space.mit.edu}

\begin{document}  
  \maketitle  
 
\begin{abstract} 
 We describe a concept for a NASA SMEX Mission in which Gas Electron
Multiplier (GEM) detectors, developed at CERN, are adapted for use in
X-ray astronomy. These detectors can be used to obtain moderately
large detector area and two-dimensional photon positions with sub mm
accuracy in the range of 1.5 to 15 keV.  We describe an application of
GEMs with xenon gas, coded mask cameras, and simple circuits for
measuring event positions and for anticoincidence rejection of
particle events. The cameras are arranged to cover most of the
celestial sphere, providing high sensitivity and throughput for a wide
variety of cosmic explosions. At longer timescales, persistent X-ray
sources would be monitored with unprecedented levels of coverage.  The
sensitivity to faint X-ray sources on a one-day timescale would be
improved by a factor of 6 over the capability of the RXTE All Sky Monitor.
\end{abstract} 
 
 
 
\section{INTRODUCTION} \label{sect:intro}
 
X-ray astronomy is our primary window to the portion of the Universe
characterized by high temperatures and explosive behavior. As one
progresses from visible light to the extreme UV, our view of the
cosmos beyond the local region in the Galaxy is increasingly
attenuated by interstellar absorption. However, as we approach 1 keV
and the X-ray band, our long-range view becomes clear again. Since
X-ray sources generally radiate fewer photons at
higher energy, it is no wonder that X-ray astronomy missions
historically favor the energy range of 1-10 keV.

Detectors with two-dimensional imaging capability are a workhorse for
X-ray astronomy.  In the energy range of 1-10 keV, there
are two primary types of position-sensitive detectors flown on
space-borne missions: gas detectors and solid-state charge-coupled
devices (CCDs).

Gas detectors include both ``standard'' proportional counters and
gas-scintillation detectors.  Position-sensitive versions of these
were used for the Einstein IPC \cite{gia79}, the ROSAT PSPC
\cite{bri88}, and the ASCA GIS \cite{koh93,oha96} for the purpose of
recording with modest spectral resolution the images formed by
focusing X-ray telescopes.  Gas detectors are also used to record the
X-ray shadows of coded masks for current wide-angle instruments,
including the All Sky Monitor\cite{lev96} of the {\em Rossi} X-ray
Timing Explorer (RXTE), the Wide Field Camera \cite{jag97} on the {\em
Beppo}SAX satellite, and the Wide-Field X-ray Monitor \cite{yam97} on
the HETE-II Mission, to be launched in mid-2000.  The dimensions of
these detectors are generally in the range of 8--25 cm, and the
position resolution is roughly 0.2--2 mm, sometimes with degraded
resolution along the second axis. Gas mixtures with argon and/or xenon
as the primary constituent provide very good quantum efficiency for
the detection of 1-10 keV X-rays, and pulse height analysis of X-ray
events provides modest spectral resolution ($\Delta E / E \sim 0.08 -
0.15$ at 6 keV).  There are a variety of methods for measuring
positions of incident X-ray photons; for the proportional counter
detectors these include charge division along resistive anodes, pulse
rise time differences, and charge collection by discrete wires, by
conductive strips on an insulating substrate, or by ``pick-up'' strips
on printed circuit boards. Both types of gas detectors routinely
provide time resolution better than 1 ms.

Recently, X-ray CCDs have been used in imaging spectroscopy missions
such as ASCA (i.e., in the SIS instrument), the Chandra X-ray
Observatory (ACIS \cite{wei00}), and Newton-XMM (EPIC). CCDs will
also be used in the Soft X-ray Cameras (SXCs) on HETE II, where they
will act as one-dimensional position-sensitive detectors in
coded-aperture cameras. Generally, CCD detectors provide superior
position resolution, in the range of 10--50 $\mu$m. CCDs
offer improved spectral resolution (e.g. $\Delta E / E \sim 0.02$ at 5
keV for the HETE-SXC). They also make it easier to extend the
sensitive energy range of the instrument down to $\sim 0.5$ keV, while
the soft X-ray sensitivity in gas detectors is usually limited by the
opacity of the entrance window (e.g. Be foil or coated polypropylene).

The use of CCDs in X-ray astronomy missions, however, has its own
limitations and difficulties. The dimension of the active area (2--6
cm) is much smaller than the size of gas detectors.  X-ray CCDs are
currently expensive, and significant resources of power, mass, and
volume are needed from the host spacecraft because of the requirements
for sophisticated drive electronics and systems for cooling the detectors
to very low temperatures. The pixel readout methods for current CCDs
provide poor time resolution (typically $\sim 1$ s) and impose
saturation limits for bright sources.  There is also susceptibility to
performance degradation in space applications, due to radiation damage
in the South Atlantic Anomaly or other high background regions.

Other types of detectors, with a range of capabilities for imaging,
such as microchannel plate detectors and microcalorimeters, also have
their place in the arsenal of detectors for X-ray astronomy, but have
characteristics which limit their applicability for instruments which
require imaging capability over a large area at reasonable cost. There
are also new materials being developed for X-ray cameras, such as
Cd-Zn-Te arrays and silicon strip detectors, which are expected to
advance our capabilities in hard X-rays, e.g. 10-100 keV. Their
limited performance below 5 keV makes it unlikely that these
technologies will compete with gas counters or CCDs in the 0.5--10 keV
band in the near future.

One may therefore conclude that there are clear tradeoffs between two
types of detectors in the current state of the art of X-ray imaging in
the main X-ray band.  Where detector area or time resolution is a high
priority, then the choice of a gas detector would have strong
advantages.  Missions in this category would include wide-field
surveys and bright-source monitors.  In the sections below, we
describe a new type of gas detector that provides the opportunity for
substantial improvements in the field of X-ray all-sky monitors. We
outline how these detectors may be adapted for astronomy, and how a
mission design for all-sky viewing would achieve major advances in
science themes related to both explosive events at short timescale and
monitoring functions at long timescales.


\section{GEM DETECTORS FOR X-RAY ASTRONOMY} \label{sect:gem}
  
In the last few years, workers at CERN have developed the Gas Electron
Multiplier (GEM) for use in gas-based particle and X-ray detectors
\cite{sau97,bach99}. The classical problem with gas counters is that a
large detector with fine anode spacing (i.e. for position resolution)
and high voltage (for soft X-ray sensitivity and good spectral
resolution) encounters substantial risk of electrical breakdown due to
the high electric fields close to the anodes.  The GEM detectors
improve this situation, essentially by working to decouple the stages
of electron multiplication and charge collection, so that both
functions are not performed by a single conductive array under high
voltage.

\begin{figure}
\begin{center}
\begin{tabular}{c}
\psfig{figure=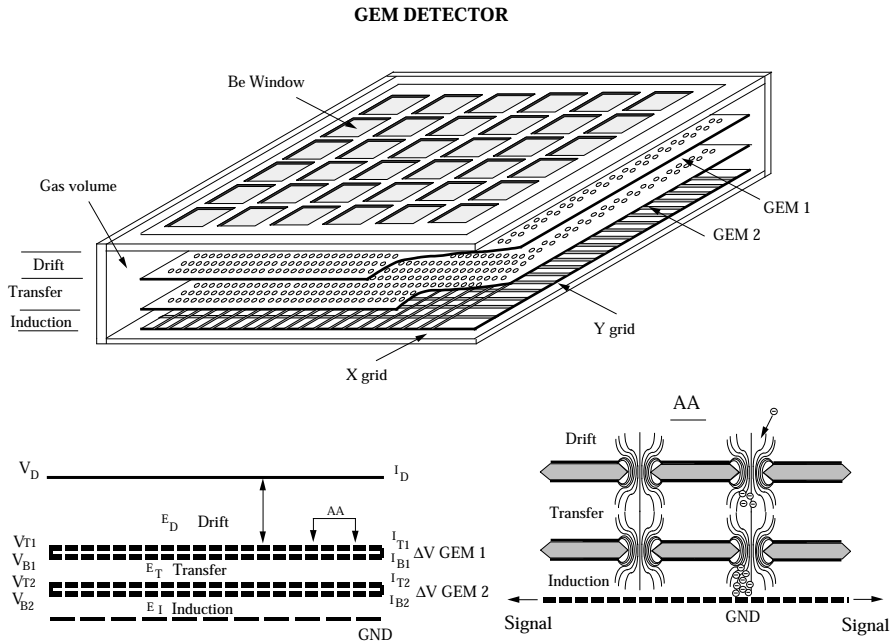,height=8cm,angle=-90} 
\end{tabular}
\end{center}
\caption[gem] { \label{fig:gem}
(top) Schematic view of a detector containing two GEM
multiplication stages.  (lower left) Schematic cross section of the
detector with the window plane at the top and readout plane at the
bottom. In this drawing, the window and detector body are at a high
negative electrical potential and the readout plane has a potential
near ground.  (lower right) A magnified view showing the electric
field configuration in and near holes in the GEM foils.
Multiplication of free electrons and ions occurs within the strong
field region inside the holes. 
}
\end{figure}

A GEM foil is a sandwich consisting of thin conductive layers of,
e.g., copper separated by a thin polymer film (e.g. 50 $\mu$m
kapton). The foil is perforated with regularly-spaced and precisely
shaped small holes (e.g., 80 $\mu$m diameter spaced with a 140 $\mu$m
pitch).  In operation inside a detector, a voltage (350--500 V) is
applied between the conductive layers.  As illustrated in Figure~\ref{fig:gem},
the electric field lines guide charged particles through the GEM
holes, with minimal losses due to collisions with the foil. Electrons are
accelerated in the strong fields within the GEM foil, producing
secondary ionizations and charge multiplication in the usual manner.

Figure~\ref{fig:gem} also depicts the operation of a GEM foil in an
X-ray detector.  When a gas atom in the upper chamber is ionized by an
incident X-ray, the initial ionization cloud drifts toward the GEM
foil due to a potential difference between the detector window and the
upper GEM layer. The electrons are vertically guided through the holes
in one or more GEM layers, with charge multiplication in the high
electric fields within each layer. The resulting electron cloud then
drifts to the readout plane, where the charge is collected via one of
several possible electrode configurations.

The GEM was initially developed for detectors that utilize a
``microstrip gas counter'' (MSGC). This detector utilizes plated
conductive strips on an insulating substrate, like a printed circuit
board, rather than a grid of self-supported wires. High voltage is
applied between the conductors in the strip.  An orthogonal strip or
an array of pickup electrodes serves to locate the event along the
second axis. MSGC detectors (without GEMs) have flown on balloon
flights \cite{ram94}, and will be flown in the JEM-X instrument
\cite{sch96} on the INTEGRAL mission (2002).

The GEM was invented to provide the charge multiplication remote from
the readout plane and to reduce the required high voltages on the
microstrip readout. In the next development, a double layer of GEM
foils was used to provide sufficient charge multiplication (e.g. gain
factors of $10^4$), so that the readout tracks could collect the
charge without operating at high voltage \cite{bre99}. GEM technology
has been successful in achieving this goal in both double-GEM and GEM
+ MSGC configurations. These detectors are successfully used outside
of CERN (e.g. \cite{van00}), and GEM + MSGC detectors are now in
``mass production'' assembly for the HERA-B experiment (CP symmetry
violation in B mesons) at the DESY accelerator in Hamburg, Germany
\cite{ste99}.

We believe that GEM-based position-sensitive proportional counters
with wireless conductive strip readout planes for charge collection
(i.e. without MSGCs) could be of great advantage for the
implementation of a number of future X-ray astronomy missions. They
may fulfill a vision of X-ray detectors with moderate cost, large
format, and excellent position resolution, while retaining the high
quantum efficiency of the noble gases for recording incident X-rays.

In moving from particle physics applications to the space environment,
several issues immediately arise.  The use of GEM detectors in space will
be most practical with sealed units in contrast to the gas-flow
detectors used in ground-based laboratories.  All of the materials
used in the detector, including the GEM foils, the GEM mounts, and the
readout planes, must not contribute contamination via out-gassing. The
detectors must be able to survive the launch and space environments
and operate reliably for many years.  Thus, the design must consider
the vibration, acoustic, and shock loads of a typical mission and also
the requirement to remain under tension and perform reliably over a
substantial range of ambient temperatures.  The detectors must also
feature readout schemes involving relatively simple electronics that
consume little power, e.g., a few watts or less per detector, in order
to be efficient and affordable for small missions.

It is highly desirable that the detector system feature a readout
plane design and electronics concept that are clearly appropriate for
SMEX and MIDEX-class missions.  One must first consider whether the
readout paradigms used in the particle physics versions of GEM
detectors are suitable in this regard.  For use in particle physics
experiments, each strip in a surface-mount readout plane is connected
to an electronic measurement chain.  This scheme requires a very large
number ($>1000$) of measurement chains.  A highly integrated readout
system for microstrip-type detectors has been developed in Europe for
use at CERN and other accelerator facilities.  The system includes the
use of a single chip, called the ``HELIX 128'', with 128 low noise
charge sensitive preamplifiers and control circuitry that can be read
out in pipeline fashion.  There is also a compatible ``SUFIX''
(SUpport chip For helIX) chip that includes an analog to digital
converter designed to interface directly with the HELIX 128 chip.  The
output from the SUFIX chip is a serial stream of digital signals
representing the charge detected in each of 128 strips or channels.
This technology is promising for future space applications of gas
detectors, but the electronics package for an instrument comprising
more than a few detectors may require a powerful CPU manager, a large
amount of power, and a relatively large volume.  While these
consequences may not be inevitable, the economic application of this
technology may not be optimal for space borne missions at the present
time.

\begin{figure}
\begin{center}
\begin{tabular}{c}
\psfig{figure=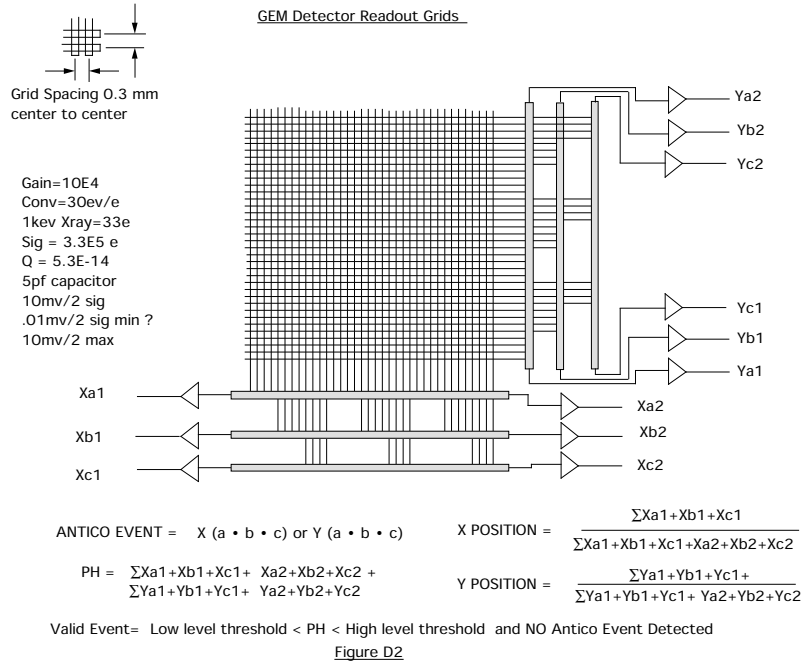,height=8cm,angle=-90} 
\end{tabular}
\end{center}
\caption[grid] { \label{fig:grid}
Schematic drawing showing the configuration of a wireless
strip readout plane.  The 6 wide bars indicate resistive
strips.
}
\end{figure}

The readout concept we intend to develop uses an array of electrodes
similar to those previously used by the developers of
the GEM \cite{bre99}.  However, rather than connecting each individual
electrode to a preamplifier and measurement chain, as is the practice
in particle physics experiments, we would connect the preamplifiers
and measurement chains to groups of anodes via resistive strips. This 
technique is analogous to the method for determining event positions
that was successfully implemented for the RXTE ASM. 
Below, we describe our approach in further detail.

We baseline a readout plane consisting of orthogonal arrays of
conductors for charge collection plated onto a substrate (see
Figure~\ref{fig:grid}).  The "X" and "Y" grids are isolated from each
other by a thin insulating layer. The X grid traces are arranged into
three groups (A, B and C) in order to help characterize the spatial
profile of the events in the detector.  Each group is made up of
multiple sets of a small number (provisionally 4) of adjacent traces;
the sets of traces are arranged in cyclic fashion, i.e., ABCABC\ldots
(see Figure~\ref{fig:grid}). The traces for each group are connected
to a resistive strip, each end of which is, in turn, connected to
readout amplifiers.  The traces forming the Y grid are also subdivided
into groups and connected to resistive strips.  The charge produced by
each event will be split among various X and Y traces and then further
split between the amplifier chains at the two ends of the associated
resistive strips. Signal charge that reaches each end of each
resistive strip will be amplified for subsequent processing including
threshold detection, background event identification, event energy and
position determination, and digitization.

\begin{figure}
\begin{center}
\begin{tabular}{c}
\psfig{figure=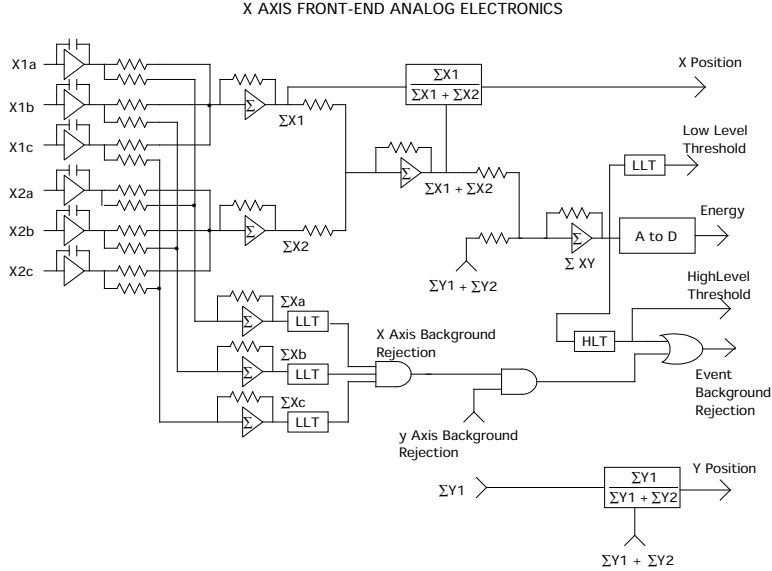,height=7cm,angle=-90} 
\end{tabular}
\end{center}
\caption[elec] { \label{fig:elec}
A simplified schematic of the readout electronics for
processing the signals from one axis of one detector.
}
\end{figure}

It is important that systems utilizing our readout and electronics
concept be able to identify and reject non-X-ray background.  We
expect that this can be accomplished in our design by two
methods. First, the total charge generated by most non-X-ray events
will exceed a high level threshold setting. Second, energetic charged
particles, as opposed to X-ray photons, will leave a long track of
secondary ionization that will normally produce an extended cloud of
electrons at the readout grid resulting in a charge signal being
produced in many of the sensing grid traces. In contrast, an X-ray
event will be more localized and produce a charge signal in only a few
of the grid traces. Background events can then be rejected by
identifying coincidences among all 3 of the A, B, and C outputs. We
expect this approach to achieve a background rejection
efficiency of 95\%. A simplified schematic of
the front-end and background detection electronics is shown in
Figure~\ref{fig:elec}.

\section{OBJECTIVES FOR AN ADVANCED X-RAY MONITOR} \label{sect:axm}

The X-ray sky is extremely variable. This variability ranges from $<$1
ms to days to years. Intensity changes by ten orders of magnitude have
been seen in some soft gamma-ray repeaters (SGRs), while even the most
persistent X-ray binaries and active galactic nuclei (AGN) commonly
vary by a factor of two or more. Space missions in the last decades
generated productive science applications related to X-ray
monitoring. Recently, X-ray source variability has been witnessed with
unprecedented clarity during the 4.5 years of observations with RXTE
(launched 12/95).  This mission includes a scanning All Sky Monitor
(ASM) and pointed instruments that provide 1 $\mu$s time
resolution. Additional insights have been gained from the
Italian-Dutch \sax Mission (launched 5/96). Scientific advances from
these observatories substantiate the perspective that high energy
astrophysics often requires a joint consideration of temporal and
spectral variability and a strong commitment to multifrequency
observations.

In a recent NASA opportunity for new missions in the Small Explorer
(SMEX) class, we proposed an ``Advanced X-ray Monitor'' (AXM) as a
concept that synthesizes the capabilities of GEM detectors and the
scientific advantages of all-sky viewing. The AXM would generate
self-contained science investigations and will further provide
invaluable and rapid guidance to other programs for scheduling
observations and for interpreting results. The AXM data and results
would constitute a public archive, and community science programs
would be funded via competitive grants administered through NASA's
Astrophysics Data Program and Astrophysics Theory Program. The AXM has
2 primary design goals:

\begin{itemize}
\item To observe nearly the entire sky, continuously, so as to
measure temporal and spectral properties of outbursts and flares 
which last from seconds to hours,
\item To monitor ``persistent'' X-ray sources with sufficient
sensitivity to provide detailed light curves for
many accreting compact objects in the Galaxy and extragalactic AGN.
\end{itemize}

The AXM concept consists of 31 cameras, each consisting of a GEM
detector mounted below a two-dimensional coded-mask aperture.  The
camera fields of view (FOV) would combine to cover the entire sky,
except for a small exclusion zone in the direction of the Sun. The
cameras would operate in the range of 1.5 to 12 keV, and they are
designed to achieve sensitivity to sources that appear for times as
short as 1 ms as well as sources that persist for many years. The
baseline AXM detection capability is $\sim300$ mCrab ($5 \sigma$) in 1
s and $\sim 1$ mCrab ($3 \sigma$) in 1~day, where 1 mCrab is $2.4
\times 10^{-11}$ erg cm$^{-2}$ s$^{-1}$ at 2-10 keV. The AXM would be
launched into a $\sim$600 km equatorial orbit. Each ``good'' event
would be telemetered to the ground with 0.3 mm spatial resolution,
$122 \mu$s temporal resolution, and 64 channels of spectral
resolution, providing enormous flexibility for scientific analyses.
The design lifetime is 3 years, with substantial probability for
longer operating lifetime.

\begin{figure}
\begin{center}
\begin{tabular}{c}
\psfig{figure=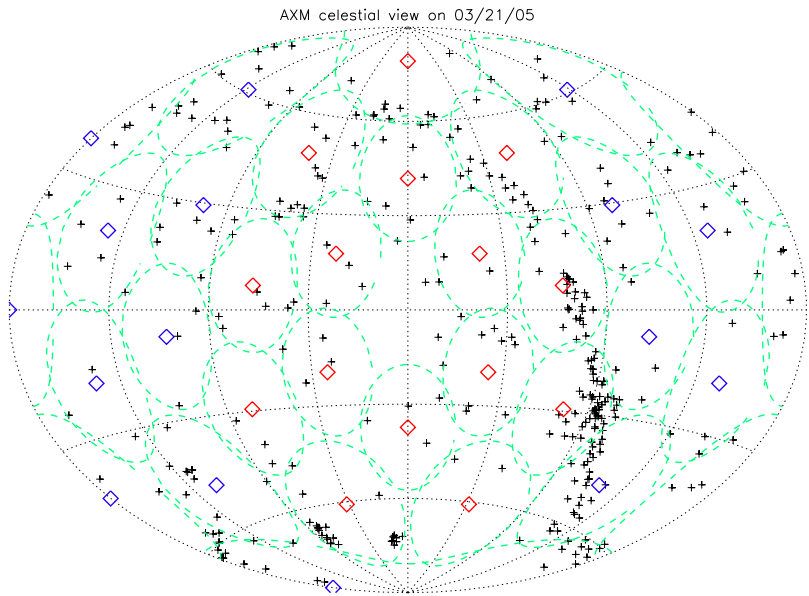,height=8cm,angle=90} 
\end{tabular}
\end{center}
\caption[sky] { \label{fig:sky} Pattern of sky coverage for the 31
cameras of the AXM when the Sun crosses the celestial equator (map
center) on the first day of spring.  The pointing directions
highlighted in red correspond to the 15 cameras oriented toward the
solar hemisphere, while the 16 positions with blue highlights view the
hemisphere away from the Sun.  The spacecraft orientation would remain
fixed except for $\sim$daily pointing adjustments to re-align the
solar axis with the Sun.  }
\end{figure}

The mounting orientations for the 31 AXM cameras are shown in
Figure~\ref{fig:sky}. They resemble the 32 faces and vertices of a
soccer ball, with 1 camera eliminated (center of figure) to avoid the
saturating effect of the Sun's X-rays. The chosen epoch for
Figure~\ref{fig:sky} is the vernal equinox; the Sun is at the center
of the figure (position 0.0, 0.0).  The 15 camera centers highlighted
in red point in the solar hemisphere, while 16 in blue view the
opposite hemisphere. The ``+'' symbols show the positions of 334
sources currently monitored with the \rxte ASM; these cluster in the
Magellanic Clouds and along the Galactic plane.  The dashed green
lines for each camera indicate offset angles of 20$^\circ$, which
corresponds to half transmission through the coded mask. The sum of
the transmission contributions produce approximately uniform exposure
over the celestial sphere, excluding only the $\sim 3$\% region near
the Sun, as well as $\sim 30$\% of the sky blocked by the Earth. A
summary of design specifications and sensitivity estimates for the AXM
are given in Table 1.

The routine on-line data analysis would consist of three
stages. First, the intensities of catalogued sources will be
determined as a function of time (for several ``standard'' time
intervals) together with confirmation or correction of the aspect of
the cameras.  Second, searches for new sources will be conducted. We
would use cross-correlation imaging techniques, e.g. at 100 s and
1-day time intervals to determine rough locations and intensities. The
data selection intervals would then be optimized to yield precise
source positions.  Third, the data analyses would then be iterated to
re-derive the intensities of previously known and new sources to
produce the final AXM results for a permanent public archive.

The results of the preliminary, on-line analysis for X-ray intensities
and new source alerts would be disseminated via the internet within
$\sim 30$ minutes of each ground station pass, in a manner similar to
what is done for RXTE ASM data.  The final results would be made
public as soon as practical, e.g. 1 week after the observations are
made.


\begin{table} [h]
\caption{A summary of AXM characteristics and performance estimates.}
\label{tab:axm} 
\begin{center}        
\begin{tabular}{|l|c|}
\hline 
\rule[-1ex]{0pt}{3.5ex} Number of cameras         &  31 \\ 
\hline 						  
\rule[-1ex]{0pt}{3.5ex} Energy Range              &  1.5--12 keV \\ 
\hline 						  
\rule[-1ex]{0pt}{3.5ex} Overall Field of View     &  97\% of $4\pi$ sr \\
\hline 						  
\rule[-1ex]{0pt}{3.5ex} Non-Earth-blocked Field   &  64\% of $4\pi$ sr \\
\hline 						  
\rule[-1ex]{0pt}{3.5ex} Operating Duty Cycle      &  95\% \\
\hline 						  
\rule[-1ex]{0pt}{3.5ex} Nominal Camera Size       &  $20\times 20\times 32$ cm \\ 
\hline 

\rule[-1ex]{0pt}{3.5ex} Spectral Resolution         &  $\sim 20$\% at 6 keV \\
\hline 
\rule[-1ex]{0pt}{3.5ex} Time Resolution             &  $122 \mu$s \\
\hline 
\rule[-1ex]{0pt}{3.5ex} Net Active Detector Area    &  200 cm$^2$ per camera \\
\hline 
\rule[-1ex]{0pt}{3.5ex} Area through the coded mask &  70 cm$^2$ per camera \\
\hline 
\rule[-1ex]{0pt}{3.5ex} Mask Area                   &  20 cm x 20 cm. \\
\hline 
\rule[-1ex]{0pt}{3.5ex} Mask Height above Detector  &  27.5 cm \\
\hline 
\rule[-1ex]{0pt}{3.5ex} Mask element (projection)   &  1 mm (12.5 arcmin) \\
\hline 
\rule[-1ex]{0pt}{3.5ex} Position bins               &  $512\times 512\times 0.3$ mm  \\
\hline 
\rule[-1ex]{0pt}{3.5ex} Uncertainty, bright sources &  1 arcmin \\ 
\hline 
\rule[-1ex]{0pt}{3.5ex} Camera 100\% transmission   &  $ 5.2^\circ$ radius \\
\hline 
\rule[-1ex]{0pt}{3.5ex} Camera 50\% transmission    &  $20.0^\circ$ radius \\
\hline 
\rule[-1ex]{0pt}{3.5ex} Total Camera FOV            &  $32.5^\circ$ radius \\ 
\hline 
\rule[-1ex]{0pt}{3.5ex} X-ray background per camera &  210 c/s \\
\hline 
\rule[-1ex]{0pt}{3.5ex} Count rate for Crab Nebula  &  170 c/s \\ 
\hline 
\rule[-1ex]{0pt}{3.5ex} Flare detection limit, 1 s  & new source: 425 mCrab \\ 
\hline 
\rule[-1ex]{0pt}{3.5ex} Flare detection limit, 1 s  & known source: 300 mCrab \\ 
\hline 
\rule[-1ex]{0pt}{3.5ex} Detection limit, 1 day      & known source: 0.8 mCrab \\ 
\hline 
\end{tabular} 
\end{center} 
\end{table} 
 

\section{AXM SCIENCE PROGRAM} \label{sect:sci}

The AXM design would address many scientific objectives 
that can be conveniently divided into three broad themes:

\begin{itemize}
\item {\bf The science of jets and other cosmic explosions}
\end{itemize}
Several types of rare, explosive events produce X-ray emission on
timescales from seconds to hours, e.g. ejections of relativistic jets
in microquasars \cite{mir99,eik98} , flares in fast X-ray novae
\cite{wij00}, flashes from SGRs \cite{gog99}, gamma ray bursts
\cite{fish95}, and giant flares from active coronae \cite{hai91}.
There is a common thread among these diverse systems: these dramatic
events are infrequent and unpredictable, and scientific progress is
impeded by the difficulty in gaining detailed observations. With
continuous viewing of 70\% of the sky and 75 cm$^2$ of active detector
area, the AXM offers substantial improvements for the capture rate of
rare, explosive events.  For comparison, the RXTE ASM cameras view
3.9\% (FWHM) of the sky with typically $\sim 20\ \rm{cm}^2$ of
effective area, and with a duty cycle of $\sim 40$\%, imposed by the
satellite orbit which traverses the SAA and other regions with high
particle flux.  The AXM also opens substantial opportunity for the
discovery of infrequent, brief outbursts of types hitherto unknown.

\begin{itemize}
\item {\bf Long term spectral and temporal evolution of persistent
sources}
\end{itemize}
Continuous all-sky viewing also provides long exposure times that may
help to address science issues that involve evolution on timescales
from several hours to several years.  In terms of both the density of
coverage and the sensitivity threshold per day, the AXM design offers
large improvements over the RXTE ASM, e.g. a six-fold increase in
signal to noise per day for faint sources, with reduced vulnerability
to systematic errors related to, e.g., the ASM's limited anode number,
time-dependent calibrations, and noisy regions of the satellite
orbit. The AXM would produce X-ray light curves which show variability
in incredible detail, e.g. tracing the spectral evolution of black
hole binaries \cite{mun99}, or tracking the excursions of accretion
flow in low-magnetic-field neutron star systems \cite{vdk00}.  Spin
changes would be measured over long baselines for {\em all}
accretion-powered pulsars to test evolving physical models for
accretion torques and disk-magnetosphere interaction
\cite{bild97,cha97}.  Pulsar spin changes in the anomalous X-ray
pulsars would test the ``magnetar'' hypothesis and the possible
association with SGRs \cite{kasp00,mel99}.

The AXM would substantially contribute to the science of Active
Galactic Nuclei (AGN) in several ways.  There would be daily
measurements of $\sim$15 BL Lac objects brighter than 1 mCrab, with
sensitivity to major eruptions (e.g. factor of 5 or more) for $\sim$80
other BL Lacs with mean flux \ga 0.3 mCrab. Since the X-ray outbursts
signify synchrotron emission that is correlated with inverse Compton
emission at much higher frequencies \cite{cat99,ghi98}, this X-ray
monitoring capability would greatly enhance the productivity of the
rapidly evolving TeV observatories (Whipple, HEGRA, CAT, etc), while
the instruments expected during the AXM era (VERITAS, HESS) are again
$\sim50 \times$ more sensitive.  For emission-line AGN, multifrequency
studies would be supported with frequent X-ray measurements of dozens of
objects in several subclasses. Furthermore, the utility of using
long-term X-ray power spectra to provide estimates for the mass of the
central black hole would be explored in earnest for the first time
\cite{ede99}.

\begin{itemize}
\item {\bf Empowerment of other observatories and multifrequency science}
\end{itemize}
All sky monitors provide alerts for new X-ray sources and for major
changes in known sources. These services provide observing
opportunities that strongly enhance the productivity of space missions
and ground based programs involved in high energy astrophysics.  In
the 2003-2007 time frame, the X- and gamma-ray missions will include
{\it Chandra}, {\it XMM}, {\it INTEGRAL}, and {\it GLAST}, and it is
extremely important that these observatories conduct their pointed
observations with full awareness of the opportunities that all-sky
viewing would provide.

There are also substantial ground based programs that are intimately
tied to the data from X-ray monitors. These include the TeV
observatories, noted above, the Greenbank Interferometer (which monitors 
high-energy sources in the radio band), and several optical programs
involving both university consortia and the growing
networks of amateur optical astronomers. There are also 
sophisticated plans for robotic observatories (e.g. Perugia, Torino,
Sarah, La Palma, Liverpool Telescope) that can contribute productive
programs for optical or IR monitoring of X-ray sources in outburst.

The advantages of combining pointed X-ray observations with all-sky
monitoring and ground-based support are abundantly clear.  Scientific
productivity is strongly enhanced by rapid response to outbursts,
state changes, etc. Alert criteria can be tailored to fit the detailed
requirements of individual research programs.  Furthermore, as shown
by the RXTE ASM, the global views of source behavior seen in long-term
X-ray light curves are widely used to provide context for, and thereby
enable more secure interpretation of, the pointed observations.

\begin{figure}
\begin{center}
\begin{tabular}{c}
\psfig{figure=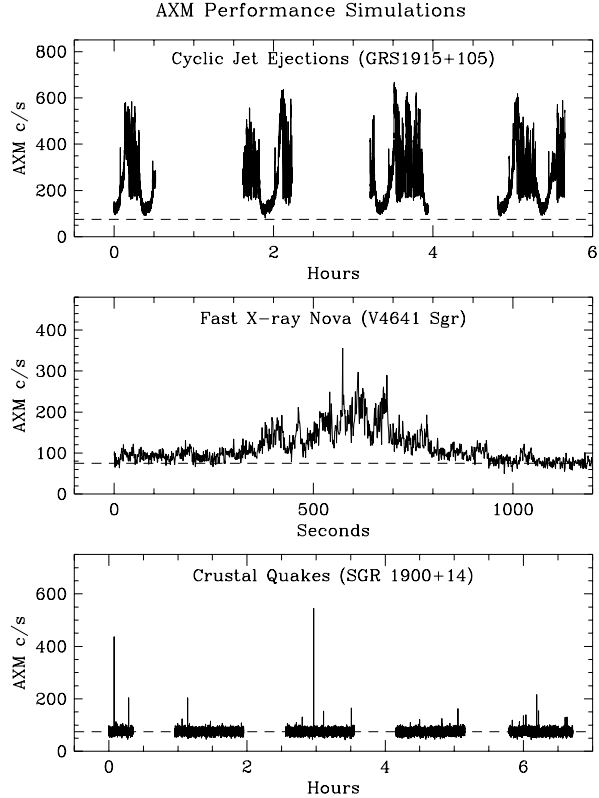,height=10cm,angle=0} 
\end{tabular}
\end{center}
\caption[sim] { \label{fig:sim} Simulations (including counting
statistics) of the performance of the proposed AXM, scaling the PCA
light curves for efffective area and adding a contribution from the
diffuse X-ray background (dashed line).  Each light
curve, shown at 1 s time resolution, is derived from the sum of
detector positions that view the source through the coded mask.
}
\end{figure}

We briefly illustrate the performance of the AXM for 3 types of explosive
X-ray sources mentioned above. We have taken the X-ray light curves
from the RXTE PCA (1 s bins) for the miscroquasar GRS1915+105
\cite{eik98}, the fast X-ray nova V4641 Sqr \cite{wij00}, and the SGR
1900+14 \cite{gog99}, and then simulated what the AXM would see from
these sources.  We scaled the PCA count rates to an effective area of
70 cm$^2$ (4.8 \% of 1 PCU), added a background count rate of 75 c/s,
and then imposed statistical noise on each bin. This is tantamount to
having a dominant source in a $20^\circ$ circle on the sky, and then
selecting only the detector positions that are open to the source (and
the diffuse background) through the coded mask. These detector
positions are generally located in two cameras. The results are shown
in Figure~\ref{fig:sim}. It is clear that the AXM contributes rich
temporal information applicable to the science of jets and other types
of cosmic explosions.

\end{document}